\documentclass[draft,aps,prd,amssymb,amsmath,eqsecnum,showpacs,nofootinbib,twocolumn] {revtex4}
\usepackage{color}
\usepackage{footnote}
\usepackage{graphicx}
\usepackage{ulem}
\usepackage{verbatim}
\usepackage{bigints}

\begin{document}
\author{ Emanuel Gallo$^1$}

\newcommand{\red}[1]{\textcolor{red}{#1}}
\newcommand{\blue}[1]{\textcolor{blue}{#1}}

\affiliation{$^1$FaMAF, UNC; Instituto de Física Enrique Gaviola (IFEG), CONICET, \\
Ciudad Universitaria, (5000) C\'ordoba, Argentina. }

\author{ Thomas M\"adler $^2$}
\affiliation{$^2$Escuela de Obras Civiles and Instituto de Estudios Astrof\'isicos, Facultad de Ingenier\'{i}a y Ciencias, Universidad Diego Portales, Avenida Ej\'{e}rcito
Libertador 441, Casilla 298-V, Santiago, Chile. }

\title{
Bounds for Lyapunov exponent of circular light orbits in black holes}

\begin{abstract}
Chaotic systems near black holes satisfy  a universal bound, $\lambda \leq \kappa_H$ linking the Lyapunov coefficient $\lambda$ associated with unstable orbits to surface gravity $\kappa_H$ of the event horizon. 
A natural question is whether this bound is satisfied by unstable circular null geodesics in the vicinity of black holes. 
However, there are known cases where this bound is violated. 
It is intriguing to ask  whether there exists an alternative universal bound that is valid in such situations. 
We show that for any spherically symmetric, static black hole that satisfies Einstein's equations and the dominant energy condition, there exist other universal bounds relating the Lyapunov coefficient to a generalized notion of surface gravity at the photon sphere.
As applications, we show how these bounds also constrain the imaginary part of quasinormal modes in the eikonal regime and how the Lyapunov coefficient relates to the shadow size and the entropy of the horizon.


\end{abstract}


\maketitle
\section{Introduction}

Lyapunov exponents, $\lambda$, quantify how rapidly initially close trajectories in phase space either converge or diverge from one another. A positive Lyapunov exponent indicates that nearby trajectories will diverge quickly, reflecting a high sensitivity to initial conditions.
In their seminal paper \cite{Maldacena:2015waa}, Maldacena and collaborators identified a universal upper bound on chaos in a thermal quantum field theory at temperature $\tilde{T}$. This bound is characterized by the Lyapunov exponent of out-of-time-ordered correlators and has in the natural units the simple expression $\lambda\leq 2\pi \tilde{T}$.
Further investigations uncovered  the validity of this universal bound in other systems, like for massive particle motions near black hole horizons \cite{Hashimoto:2016dfz}, a connection to the presence of a horizon \cite{Dalui:2018qqv}, and how potential quantum corrections might enhance chaotic behavior of massless particles near the horizon \cite{Bera:2021lgw}.
Indeed, many of the investigated systems  satisfy the inequality
\begin{equation}
    \lambda \leq \kappa_H ={2\pi T_H},\label{eq:1lambda}
\end{equation}
{where $\kappa_H$ is the 
surface gravity of the horizon of the black hole and $T_H$ its associated temperature.} For other studies were this bound is satisfied or violated in different spacetimes and gravitational theories we refer to \cite{Wang:2023crk,Lei:2024qpu,Tang:2023wyn,Addazi:2021pty,Addazi:2023pfx,Gao:2022ybw,Hashimoto:2022kfv,Yu:2023spr,Xie:2023tjc, Kan:2021blg,Gwak:2022xje,Park:2023lfc}. Alternative methods to study chaos can be found in \cite{DeFalco:2020yys,DeFalco:2021uak}.

Circular orbits of null geodesics play a role in upcoming space-based very long baseline interferometry missions. 
The flux ratio between consecutive light rings can be related to the Lyapunov exponents of corresponding adjacent nearly bound null geodesics \cite{Johnson:2019ljv}. 
Thus an insight into the Lyapunov exponents of these orbits offers a better understanding of black hole observations and thereby provides means to test modified theories of gravity. 
In fact, several authors analyzed such dependence for various spacetime geometries with the aim of placing constraints on alternative theories through long-baseline observations \cite{Staelens:2023jgr, Deich:2023oox, Salehi:2024cim, Gralla:2019drh, Kocherlakota:2024hyq, Wong:2024gph}.

The Lyapunov exponent associated with unstable circular-photon null geodesics is also related to lensing in the strong deflection limit of black holes \cite{Stefanov:2010xz,Raffaelli:2014ola}, to the imaginary part of quasinormal modes in the eikonal limit \cite{Cardoso2009}, and to the phase-transition properties of black holes which multiple branches correspond to the distinct phases of the black hole \cite{Guo:2022kio, Kumara:2024obd, Shukla:2024tkw}.

From the observational importance of the Lyapunov exponent {arises the   natural question: Does relation \eqref{eq:1lambda} hold for unstable  null circular geodesics? 
As a matter of fact for many spacetimes the answer is positive, but a corresponding relation for near-extremal black holes does not hold \cite{Yu:2022tlr, Bianchi:2020des}. 
The reason for the latter is that in the limit towards extremality the surface gravity $\kappa_H$ approaches zero, but the Lyapunov exponent does not.
Consequently, there is a region in the parameter space  of nearly extremal black holes where the  bound is violated.

 Nevertheless,  \cite{Abreu:2010sc} introduced a concept of a ``generalized surface gravity", $\kappa(r)$, for static, spherically symmetric spacetimes, which is applicable at arbitrary radial positions\footnote{{ In \cite{Abreu:2010sc}, the concept of generalized surface gravity was introduced in the context of a general static spacetime. For our purposes, we limit their study to static and spherically symmetric black holes.}} $r$.
 In particular, evaluation of $\kappa(r)$ at the position $r_H$ of a  black hole horizon yields its surface gravity $\kappa(r_H) = \kappa_H$. {As shown in \cite{Abreu:2010sc}  the usage of $\kappa(r)$  provides simple bounds on quasi-local entropies  similar to the holographic bounds relating the entropy $S(r)$ with area $A(r)$ of the enclosed region, namely, \begin{equation}
     S(r)\leq \frac{\kappa(r)}{\kappa_H}\frac{A(r)}{2},
     \end{equation}
     with $\kappa_H$ the standard surface gravity associated to the event horizon.}
     
 Moreover, at the radial position $r_\gamma$ of  unstable circular null geodesics, $\kappa(r_\gamma)$ is non-vanishing even for nearly extremal black holes. 
 This provokes the question if a similar relation to eq. \eqref{eq:1lambda} holds when replacing $\kappa_H$ by the generalized surface gravity $\kappa_\gamma\equiv \kappa(r_\gamma)$. That is, setting a bound to the Lyapunov exponent of circular null geodesics with the value of the corresponding generalized surface gravity at $r_\gamma$. 

Our answer to this question is affirmative. 
For {\it any} spherically symmetric black hole satisfying the Einstein equations and the dominant energy condition, there exists a universal upper bound relating $\lambda$ and $\kappa_\gamma$.
This upper bound also explicitly  depends  on the radial position of the photon sphere $r_\gamma$.
Furthermore,  we show the existence of a bound 
between the Lyapunov exponent and  {an operational} Unruh temperature $T_U=\frac{a_U}{2\pi}$  { with $a_U$ being the norm of the proper acceleration of} static observers located at $r_\gamma$ . {We refer to this temperature as operational because, in reality, a static observer near a black hole would measure a different temperature, namely the Hawking radiation temperature adjusted by a redshift factor that depends on the observer's location. However, this operational temperature emerges in effective geometries associated with the photon sphere \cite{Raffaelli:2021gzh,Giataganas:2024hil}. To facilitate comparison with those works we have chosen to express some of the derived inequalities in terms of $T_U$. Notably, in \cite{Barbado:2016nfy}, the authors demonstrated that the temperature $T_U$ is indeed measured by radially free-falling observers starting from rest at a radius $r_\gamma$ for radiation in the so-called ingoing sector\footnote{{While \cite{Barbado:2016nfy} computations are valid for any fixed radius $r$, we specifically refer to $r_\gamma$ as our discussion is confined to the photon sphere region.}} (i.e., radiation pointing towards the black hole). That is, while a static observer with acceleration $a_U$ detects Hawking radiation, a radially free-falling observer instantaneously at rest there measures ingoing radiation with a temperature $T_U$ (at the initial time, when the observer is at rest\footnote{{These observers also measures outgoing radiation with contribution of Hawking and Unruh effect. For other positions of the radially free-falling observer, the temperature of the ingoing radiation has Doppler corrections due to the relative velocity between the free-falling observer and static observers (see \cite{Barbado:2016nfy} and references therein for more details).}}), which precisely matches $\frac{a_U}{2\pi}$.} The  bounding relation between Lyapunov exponent and operational Unruh temperute,
$\lambda \leq 2\pi T_U$, is saturated for a Schwarzschild black hole. 
In fact, it was recently derived in  \cite{Giataganas:2024hil} for the two particular cases of Schwarzschild and Reissner-Nordström black holes. 
Our proof generalizes their results to a broader class of spherically symmetric black holes and is sufficiently general to not require an explicit form of the metric.
Moreover, our new bound is directly related to recent bounds associated with the shadow size and acceleration bounds for radial linear uniformly accelerated  trajectories in static spherically symmetric black hole spacetimes of the Schwarzschild type \cite{Paithankar:2023ofw}.

Our major results culminate in the expressions for the generic bounds between the Lyapunov exponent and 
(i) surface gravity at the photons sphere (eq. \eqref{lambdaboundkappa},  \eqref{eq:lambda20} and \eqref{eq:lambdar+}), (ii) the acceleration of static observers at the photon sphere (eq. \eqref{eq:nueva}), and (iii) the shadow of the black hole (eq. \eqref{eq:lambdars}). 
 
As an application of these new bounds, we derive a constraint for the quasinormal modes under the assumption of validity of the Wentzel-Kramers-Brillouin (WKB) approximation\cite{1985ApJ291L33S} in the eikonal regime. Following the approach in \cite{Giataganas:2024hil}, we also present simple bounds for the Lyapunov exponent using the inverse of the entropy of the black hole horizon.

The note is organized as follows: in Sec.~\ref{sec:BG} we outline the general geometric setup; in Sec.~\ref{sec:LE}, we derived the different expression for the bounds of the Lyapunov exponents; and in Sec.~\ref{sec:FIN}, the major results as summarized and possible future work is indicated. Natural units $G=c=h=1$ are used throughout and the metric signature is $+2$.

\section{Background}\label{sec:BG}
Consider a 4-dimensional, spherically symmetric and static spacetime with a metric described by 
\begin{equation}\label{eq:metric}
\begin{split}
ds^2 =& -
\frac{\mu(r)}{e^{2\delta(r)}}dt^2 + \frac{dr^2}{\mu(r)}
+ r^2 \left(d\theta^2 + \sin^{2}\theta \, d\phi^2\right),
\end{split}
\end{equation}
We  assume that the spacetime contains an  event horizon of a black hole at $r=r_H>0$.  
We have $\mu(r_H)=0$ at the event horizon and  require the additional conditions \cite{Nunez:1996xv}
\begin{equation}\label{eq:condition}
\mu^\prime(r_H)\ge0\;\;,
\delta(r_H)\le\infty\;\;,\;\;
\delta^\prime(r_H)\le\infty\;\;.\;\;
\end{equation}

In the exterior $r>r_H$, we demand the  asymptotical flatness conditions
\begin{equation}
\lim_{r\rightarrow\infty}\mu=1\;\;,\;\;
\lim_{r\rightarrow\infty}\delta=0\;\;.\;\;\label{eq:deltaprime}
\end{equation}

Einstein equations are $G_{ab} = 8\pi T_{ab}$ with the Einstein tensor $G_{ab}$ and  energy-momentum tensor $T_{ab}$ {with corresponding trace $T:=g^{ab}T_{ab}$}.
The energy density as measured by static observers is given by $\rho=-T_{ab}t^at^b $  with the timelike unit vector $t^a$ which  is proportional to the timelike Killing vector field $\partial_t$.
Based on  $t^a$ we form the orthonormal tetrad ${E_\alpha = (t^a, e^a_{(r)}, e^a_{(\theta)},e^a_{(\phi)})}$. 
This tetrad defines the radial pressure $p_r = T_{ab}e^a_{(r)}e^b_{(r)}$, and the pressure $p_\bot$ tangent to the orbits of spherical symmetry, i.e.  $p_\bot=T_{ab}e^a_{(\theta)}e^b_{(\theta)} =T_{ab}e^a_{(\phi)}e^b_{(\phi)}$. Consequently, the trace of the energy momentum tenor is ${T = -\rho+p_r+2p_\bot}$. We require the  dominant energy condition \cite{Hawking:1973uf,Wald:1984rg}, $\rho\ge0$, $\rho\ge|p_r|$ and $\rho\ge|p_\bot|$. 

The function $\mu$ relates to the Misner-Sharp mass $m(r)$ \cite{Misner:1964je}
\begin{equation}\label{MSmass}
m(r) = \frac{r}{2} (1-\mu) = m_H+4\pi\int_{r_H}^r \rho r^2dr,
\end{equation}
where we defined the horizon mass $m_H$. The  ADM mass $M$ is the obtained from the limit of $m$ as $r\to\infty$.  Finiteness of the limit of $m$ implies by \eqref{MSmass} that 
\begin{equation}\label{eq:limitrho}
\lim_{r\rightarrow \infty} r^3\rho = 0\;\;.
\end{equation}

From the $(tt)$ and $(rr)$ components of field equations  and from $\nabla_a T^a_r= 0$ we get \cite[for $\hat\alpha=0$ and $n=4$]{Gallo:2015bda}
\begin{align}
8\pi\rho=&
=\frac{1}{r^2} - \frac{(r\mu)^\prime}{r^2}\;,\label{EE_rho}\\
8\pi p_r=&\frac{e^{2\delta}}{r^2}\left(\frac{r\mu}{e^{2\delta}}  \right)^\prime -\frac{1}{r^2}
\;,\label{EE_pr}\\
p'_r=&-\frac{(e^{-2\delta}\mu)'}{2e^{-2\delta}\mu}(\rho+p_r)+\frac{2}{r}(p_\bot-p_r)\;, \label{EE_dT}
\end{align}
where $^\prime$ denotes the derivative with respect to $r$. The sum of \eqref{EE_rho} and \eqref{EE_pr} and evaluation at the horizon gives with $\mu(r_H)=0$ that $\rho(r_H) +p_r(r_H) = 0$.
Based on eqs.\eqref{EE_rho} and \eqref{EE_pr}, we have
\begin{equation}\label{eq:delta}
\mu\delta' = -4\pi r(\rho + p_r).
\end{equation}

For any  metric of the form \eqref{eq:metric}, the radial coordinate of null geodesics satisfies an equation of the form $\dot{r}^2 = V(r,E,L),$
where a dot denotes a derivative with respect to an affine parameter, and $V$ is an effective potential given by \cite{Cardoso2009}
\begin{equation}\label{eq:effectiveV}
    V=e^{2\delta(r)}E^2-\frac{L^2}{r^2}\mu(r),
\end{equation}
with $E$ the energy and $L$ the orbital angular momentum of the massless particle. 
Circular null geodesics at ${r=r_\gamma}$ satisfy $V = V' = 0$. {For} the metric (\ref{eq:metric}), it is equivalently to require 
$\tilde{N}(r_\gamma)=0$ (see for example
\cite{Cardoso2009,Hod:2013jhd})
with
\begin{equation}\label{eq:ntilde}
\tilde{N}(r)=2e^{-2\delta(r)}\mu(r)-r[e^{-2\delta(r)}\mu(r)]'.
\end{equation}
Alternatively,  circular null geodesics can be  obtained from the vanishing of the function $N(r)=e^{2\delta(r)}\tilde{N}(r)$,
which
by using   \eqref{EE_pr} can be expressed as
\begin{equation}
N(r)=3\mu(r)-1-8\pi r^2p_r.\label{NGR}
\end{equation}
For circular null geodesics we also have $E=\pm\frac{\sqrt{e^{-2\delta(r_\gamma)}\mu(r_\gamma)}}{r_\gamma}L.$

As first shown by Hod \cite{Hod:2013jhd}, eq.~\eqref{NGR} admits at least one solution, which follows from the following facts: 
First, the conditions $\mu(r_H) = 0$ and \eqref{eq:condition} imply that $N(r)$ satisfies $N(r_H) \leq 0$. 
Second, taking into account that $\lim\limits_{r \rightarrow \infty} r^2 p_r = 0$,  which follows from the dominant energy condition and \eqref{eq:limitrho}, we see that $N(r \to \infty) \to 2$. 
Therefore, there exists a value $r_\gamma$ such that $N(r_\gamma) = 0$.
Since we are interested in the innermost null circular orbit defined by $r = r_\gamma$, $N(r)$ satisfies
\begin{equation}\label{N22}
N(r_H \leq r < r_\gamma) < 0,
\end{equation}
and
\begin{equation}\label{N3}
N'(r_\gamma) \geq 0.
\end{equation}
The stability of the orbits follows from considering the second derivative of the effective potential $V$ given by \eqref{eq:effectiveV}, whose evaluation at $r_\gamma$ gives
\begin{equation}
V''({r_\gamma}) = \frac{L^2}{r^4 e^{-2\delta{(r)}}}\left[2e^{-2\delta{(r)}} \mu(r) - r^2_\gamma [e^{-2\delta(r)}\mu(r)]''\right] \bigg|_{r_\gamma},
\end{equation}
which, using the Einstein equations \eqref{EE_rho}-\eqref{EE_dT}, can be rewritten as \cite{Gallo:2015bda},
\begin{equation}\label{eq:veeff}
V''(r_\gamma) = \frac{L^2 N'(r_\gamma)}{r_\gamma^3}.
\end{equation}
Therefore, taking into account \eqref{N3}, we obtain $V''(r_\gamma) \geq 0$, implying that  in general $r_\gamma$ corresponds to an unstable circular orbit\footnote{Note that the equation $N(r)=0$ could have more than one solution. From its behavior at $r_H$ and at the asymptotic region the outer light ring must be unstable. 
However, here we are interested in the inner circular orbit where $N'(r_\gamma)\geq 0$. 
If $N'(r_\gamma)>0$, the instability follows. 
But for the special case of $N'(r_\gamma)=0$, the stability behavior would depend on the behavior of higher derivatives of $N(r)$. 
However, the case $N'(r_\gamma)=0$ is only possible for a very special case where the matter content is such that $ (\rho(r_\gamma) + p_\bot(r_\gamma))= (8\pi r^2_\gamma)^{-1}$ (see Eq.\eqref{eq:Nprime}).
}.

Since most matter fields obey $T\le0$, Hod \cite{Hod:2013jhd} used this condition to prove that $\mu(r_\gamma)$ obeys

\begin{equation}\label{eq:mu13}
\mu(r_\gamma)\leq\frac{1}{3},
\end{equation}
which  by using \eqref{MSmass} gives
\begin{equation}\label{eq:hod2}
r_\gamma\leq 3{m(r_\gamma)}\leq 3{M}.
\end{equation}
This bound was extended to higher dimensions and various gravitational theories \cite{Gallo:2015bda, Cvetic:2016bxi, Ma:2019ybz}, and to other compact objects \cite{Peng_2020, Xavier:2024iwr}. Compact objects, including black holes with multiple photon rings, were discussed in\cite{Cunha:2017qtt, DiFilippo:2024ddg, Xavier:2024iwr, Hod:2017zpi, Guo:2022ghl}, and \cite{Hod:2022mys, Peng:2022edz} show that extremal black holes generally have external light rings.

\section{Bounds on the Lyapunov exponent of circular null geodesics}\label{sec:LE}
 For massive particle motion near the horizon of a black hole, there exists a relation between the Lyapunov exponent $\lambda$ of their phase space trajectories and the surface gravity $\kappa_H$ of the event horizon, specifically, $\lambda \leq \kappa_H.$

Hereafter, we briefly summarize the computation of the Lyapunov exponents for the motion of test particles, following \cite{Cornish:2003ig, Cardoso2009}. The equations of motion of $N$ particles can be schematically be written as 
\begin{equation}
\frac{dX_i}{dt} =  H_i(X_j, g_{ab}),
\end{equation}
where $X_i$ is a 2N dimensional vector consisting of the position and velocities (or momenta) of the particles and $H_i(X_j)$  is a $2N$ dimensional vector  depending on $X_i$, the metric and the metric's first derivatives.  
Linearisation around an  orbit $X_i$ yields
\begin{equation}\label{eq:ddxdt}
\frac{d \delta X_i(t)}{dt} = K_{ij}(t) \delta X_j(t),
\end{equation}
where $K_{ij}(t) = \frac{\partial H_i}{\partial X_j}\bigg|_{X_i(t)}$ is the stability matrix. 
A solution to \eqref{eq:ddxdt} is
\begin{equation}
\delta X_i(t) = L_{ij}(t) \delta X_j(0),
\end{equation}
with $\dot{L}_{ij}(t) = K_{im} L_{mj}(t)$ and $L_{ij}(0) = \delta_{ij}$. 

The Lyapunov exponent $\lambda$ is determined by 
\begin{equation}
\lambda = \lim_{t \to \infty} \frac{1}{t} \log \left[\frac{L_{jj}(t)}{L_{jj}(0)} \right].
\end{equation}
For one particle with a two-dimensional phase space $X_i(t) = (p_r, r)$, such as circular orbits in spherically symmetric spacetimes, the equations linearize to
\begin{equation}
K_{ij} = \begin{pmatrix} 0 & K_1 \\ K_2 & 0 \end{pmatrix}.
\end{equation}
with stability matrix components given by 
\begin{eqnarray}
K_1 &=& \frac{d}{dr} \left( \dot{t}^{-1} \frac{\partial \mathcal{L}}{\partial r} \right),\\
K_2 &=& - \dot{t} g_{rr}^{-1},
\end{eqnarray}
where $\mathcal{L}=\frac{1}{2}g_{\alpha\beta}\dot{x}^\alpha\dot{x}^\beta$ the Lagrangian for geodesic motion. 
The principal Lyapunov exponents for circular orbits can then be expressed as
\begin{equation}
\lambda^2 = K_1 K_2.
\end{equation}
In particular, the Lyapunov exponent for circular null geodesics of metrics of the form \eqref{eq:metric} is given by \cite{Cardoso2009} 
\begin{equation}
\lambda = \sqrt{\frac{r_\gamma^2 [e^{-2\delta(r_\gamma)} \mu(r_\gamma)]}
{2L^2} V''(r_\gamma)}, \label{lambda}
\end{equation}
which while using \eqref{eq:veeff} yields  \cite{Gallo:2015bda}
\begin{equation}\label{eq:lyap}
\lambda = \sqrt{\frac{e^{-2\delta} \mu}{2r} N'} \bigg|_{r_\gamma},
\end{equation}
which is a real quantity due to eq.~\eqref{N3}.

For null geodesics at the photon sphere, the bound \eqref{eq:1lambda} can be violated where $\lambda$ is computed via \eqref{eq:lyap}. 
Examples are near-extremal Reissner-Nordström black holes with a charge $Q\in [0.99M, M]$ as shown in  \cite{Bianchi:2020des}.

To establish a bound relating the Lyapunov exponent with a surface gravity, which does not exclude these cases, we employ the notion of the generalized surface  gravity $\kappa(r)$  of \cite{Abreu:2010sc} which is valid for every sphere of radius r. {{It is important to clarify that this concept of surface gravity does not imply the existence of a quasilocal version of the first law of thermodynamics at fixed radii $r_o$, relating changes in energy $\delta E$ of infalling objects crossing the timelike surface $r = r_o$ (as measured by static observers at that location) to changes in the area $\delta A$ of the black hole's event horizon through $\delta E = \frac{{\kappa(r_o)}}{8\pi} \delta A$. Although a local version of this law  exists in the literature \cite{Frodden:2011eb,Tripathy:2023mpi}, it refers to an effective surface gravity given by $\tilde{\kappa} = \frac{\kappa_H}{|t^a|_{r_o}}$, with $|t^a|_{r_o}$ the norm of the timelike Killing vector $t^a$ at $r = r_o$. However, as shown in \cite{Ravuri:2024fva},  $\kappa(r)$ contributes to local temperatures of radially free-falling observers. A generalization of $\kappa(r)$ for non-static spacetimes can be found in \cite{Barcelo:2010xk}.}} 
This generalized surface gravity $\kappa(r)$ is
determined by a redshift factor of the norm $a = (g_{bc}a^ba^c)^{1/2} $ of the acceleration $a^b=t^c\nabla_ct^ b$ of observers following the integral curves along  the unit vector field $t^c$. For \eqref{eq:metric}, {it} results in 
\begin{equation}
    \kappa(r) = \sqrt{e^{-2\delta} \mu} a(r),
\end{equation}
with 
\begin{equation}\label{eq:acc}
a(r) = \frac{m(r) - r m'(r)}{r^2\sqrt{\mu}}- {\delta'(r)}{\sqrt{\mu}}.
\end{equation}
Using the Einstein equations in \eqref{eq:acc}, we obtain
\begin{equation}
a(r) = \frac{1}{\sqrt{\mu}} \left( \frac{m(r)}{r^2} + 4\pi r p_r \right),
\end{equation}
which allows us to rewrite $\kappa(r)$ as

\begin{equation}\label{eq:kappaViss}
    \kappa(r) = \frac{e^{-\delta}}{r^2}\left[m(r) + 4\pi r^3 p_r\right]  
    =\frac{e^{-\delta}}{r}\left(\mu - \frac{N}{2}\right),
\end{equation}
where we have expressed $m(r)$ in terms of $\mu$, and $p_r$ in terms of $N$ using Eq.\eqref{NGR}. 
 In most physical situations, one would expect  $\kappa(r) < \kappa_H$ in the exterior region of a black hole horizon \cite{Abreu:2010sc}. 
 However, for near-extremal black holes, where $\kappa_H \to 0$, the generalized surface gravity does remains positive for $r\rightarrow r_\gamma$, meaning $\kappa_\gamma=\kappa(r_\gamma)$ does not vanish. 
 Therefore, it is natural to seek an upper bound for the Lyapunov exponent in terms of $\kappa_\gamma$ instead of $\kappa_H$. 

Evaluation of \eqref{eq:kappaViss} at the photon radius $r_\gamma$ and using  $N(r_\gamma)=0$, we find

\begin{equation}\label{eq:fundam}
\kappa_\gamma = \frac{e^{-\delta}\mu}{r}\bigg|_{r_\gamma}= \sqrt{e^{-2\delta(r_\gamma)}{\mu(r_\gamma)}}a_{U},
\end{equation}
where $a_{U} = a(r_\gamma)$.
This  norm of the acceleration  at $r_\gamma$ relates to $T_{U}(r_\gamma)$ by
\begin{equation}
a_{U}=\frac{\sqrt{\mu}}{r}\bigg|_{r_\gamma}=2\pi T_{U}(r_\gamma).\label{eq:unruh2}
\end{equation}

To relate \eqref{eq:lyap} with \eqref{eq:kappaViss}, we need to calculate the derivative of $N$ which will contain the derivative of $p_r$. For the latter, it follows from the vanishing of Eq.\eqref{eq:ntilde} and Eq.\eqref{EE_dT} that
\begin{equation}\label{eq:dprdr}
    p'_r(r_\gamma)=\frac{1}{r}(2p_\bot-3\rho+p_r)\bigg|_{r_\gamma}.
\end{equation}
Thus, by differentiating Eq.\eqref{NGR} with respect to $r$ and taking into account Eq.\eqref{EE_rho} {and \eqref{eq:dprdr} }together with the dominant energy condition, {i.e. $\rho + p_\bot \geq 0$,} we find
at $r_\gamma$
\begin{equation}\label{eq:Nprime}
N'(r_\gamma) = \frac{2}{r}\left[1 - 8\pi r^2 (\rho + p_\bot)\right]\bigg|_{r_\gamma}\leq \frac{2}{r_\gamma}.
\end{equation}
Therefore, we   are in condition to prove  the following inequality:
\begin{equation}\label{lambdaboundkappa}
\begin{split}
\lambda=&\sqrt{\frac{e^{-2\delta}\mu}{2r}N'}\bigg|_{r_\gamma}\\
=&\frac{\kappa_\gamma}{\sqrt{\mu}_\gamma} \left({1-8\pi r^2 (  \rho_{\gamma}+p_{\bot\gamma})}\right)^{1/2}
\leq \frac{\kappa_\gamma}{\sqrt{\mu_\gamma}},
\end{split}
\end{equation}
relating $\lambda$ to $\kappa_\gamma$, but also depending on $r_\gamma$ through $\mu_\gamma$. Since in general at $r_\gamma$ we only know that $\mu_\gamma \leq \frac{1}{3}$ (assuming a no positive trace of $T^{ab}$), we see that the inequality $\lambda \leq \kappa_\gamma$ cannot be guaranteed by eq.\eqref{lambdaboundkappa}. 
In fact, for an extremal Reissner-Nordström metric, we have $\lambda = \frac{\sqrt{2}}{8}$, while $\kappa_\gamma = \frac{1}{8}$, and therefore, $\lambda > \kappa_\gamma$. Nevertheless, in consistence with our proof, it can be checked that the bound in eq. \eqref{lambdaboundkappa} is satisfied for any Reissner-Nordström black hole, and in particular for an extremal Reissner-Nordström black hole where at $r_\gamma$, $\mu_\gamma = \frac{1}{4}$\footnote{For a RN black hole with ADM mass $M$ and charge $Q$, $m(r)=M-\frac{Q^2}{2r}$ and the outer photon sphere is placed at $r_\gamma=\frac{3}{2}M+\frac{1}{2}\sqrt{9M^2-8Q^2}$. 
In particular for an extremal Reissner-Nordström black hole, $r_\gamma=2M$, and $m(r_\gamma)=3M/4$.}.

 Alternatively, we can express \eqref{lambdaboundkappa} explicitly in terms of $\kappa_\gamma$ and $r_\gamma$. 
 This follows from  \eqref{eq:deltaprime} and \eqref{eq:delta} and the dominant energy condition, which imply $\delta' \leq 0$ and therfore $e^{-\delta(r_\gamma)} \leq 1$. 
 Under these circumstances, we get from  \eqref{eq:lyap} and \eqref{eq:kappaViss}
\begin{equation}\label{eq:lambda20}
\begin{split}
    \lambda&=\sqrt{\frac{e^{-2\delta}\mu}{2r}N'}\bigg|_{r_\gamma}=e^{-\delta}\sqrt{\frac{e^{-\delta}\mu}{r}\frac{{1-8\pi r^2 (  \rho_{\gamma}+p_{\bot\gamma})}}{r}}\bigg|_{r_\gamma}\\
&\leq e^{-\delta}\sqrt{\frac{\kappa}{r}}\bigg|_{r_\gamma}\leq \sqrt{\frac{\kappa_\gamma}{r_\gamma}}.
\end{split}
\end{equation} 
The bound \eqref{eq:lambda20} also implies  $\lambda^2 r_\gamma \leq {\kappa_\gamma}$.
{Since $r_\gamma > r_H$,} we obtain the simple inequality 
\begin{equation}\label{eq:lambdar+}
\lambda^2 r_H \leq \kappa_\gamma.
\end{equation}

It results that $\lambda$ is bounded by $\kappa_\gamma$ and for the size of the event horizon. 
In fact, this inequality can be improved for black holes with a traceless energy-momentum tensor. 
In such cases, Hod recently showed in \cite{Hod:2023jmx} that $r_\gamma \geq \frac{6}{5}r_H$, which implies that $\lambda^2 r_H \leq \frac{5}{6}\kappa_\gamma$.

 It turns out that the relations \eqref{lambdaboundkappa}, \eqref{eq:lambda20} and \eqref{eq:lambdar+} are not the only inequalities which can be written at the photon sphere of  spherically symmetric black hole under the given physical assumption.  In fact, we can re-express the inequality \eqref{lambdaboundkappa} in a very natural way in terms of the ``Unruh temperature'' $T_U$. To do so, note that from eq. \eqref{eq:fundam} and the dominant energy condition, it follows that
\begin{equation}\label{eq:kaubu}
    \frac{\kappa}{\sqrt{\mu}}\bigg|_{r_\gamma}=e^{-\delta(r_\gamma)}a_{{U}}\leq a_{{U}},
\end{equation}
which by using \eqref{eq:unruh2} and \eqref{lambdaboundkappa} implies the universal upper bound for the Lyapunov exponent:
\begin{equation}\label{eq:nueva}
    \lambda\leq a_U= 2\pi T_{U}.
\end{equation}
It is worth noting that, recently, in reference \cite{Giataganas:2024hil}, the validity of this inequality was observed for the special cases of Schwarzschild and Reissner-Nordström black holes through direct calculation. 
What we have shown in Eq.\eqref{eq:nueva} is that it remains valid within Einstein's theory of general relativity, under very natural physical assumptions.
 Note also, that a relation between the Lyapunov exponent and the Unruh temperature associated to a scalar field thermalization in the vicinity of an effective metric near to the photon sphere was also observed in \cite{Raffaelli:2021gzh}. 
 
 In turn, \cite{Paithankar:2023ofw} showed that metrics like \eqref{eq:metric} but with $\delta(r) = 0$, there exists an equality between the size of the black hole shadow and the acceleration $a_b$ associated with uniformly accelerated radial motion, which passes through $r_\gamma$. Specifically, they found that the radius of the shadow $R_{sh}$ is given by $R_{sh} = a_b^{-1}$. 
 Moreover, it follows that  $a_b = a_U$ so that $R_{sh} = a_U^{-1}$.
 As the most general radius $R_{sh}$ of the shadow is given by
\begin{equation}\label{eq:Rsh_au}
    R_{sh} = \frac{r_\gamma}{\sqrt{e^{-2\delta(r_\gamma)}\mu(r_\gamma)}} = \frac{e^{\delta(r_\gamma)}}{a_U}.
\end{equation}
We can see that we have a relation between the Unruh acceleration and the shadow radius, namely $a_U = e^{\delta(r_\gamma)}R_{sh}^{-1}$,
which provides a link between the Lyapunov exponent and the shadow size of the black hole using \eqref{lambdaboundkappa}, \eqref{eq:kaubu}  and \eqref{eq:Rsh_au},  
\begin{equation}\label{eq:lambdars}
    \lambda \leq \frac{1}{R_{sh}}.
\end{equation}

This inequality has already appeared in \cite{Bianchi:2020des} for the special family of spherically symmetric and charged Anti-de-Sitter metrics, where it was verified to be valid by explicit computation. In that reference, the inequality is expressed in terms of the impact parameter $b$ (which for an asymptotic observer coincides with $R_{sh}$). Once again, our results generalize this inequality for a broader family of black holes.

As final remarks,  the  bounds \eqref{lambdaboundkappa}, \eqref{eq:lambda20}, and \eqref{eq:lambdar+} together with the assumption of a non-positive trace of the energy momentum tensor imply  new (weaker) bounds relating $\lambda$ directly with $r_\gamma$ or $r_H$. 
Indeed, \eqref{eq:mu13},   \eqref{eq:fundam}, the dominant energy condition and the Einstein equations imply that ${\kappa_\gamma r_\gamma \leq \frac{1}{3}}$. Hence, as $r_\gamma>r_H$  we have from \eqref{eq:lambdar+}
\begin{equation}\label{eq:lyapuni}
\lambda\leq\frac{1}{\sqrt{3}r_H}.
\end{equation}
This inequality allows to write a bound for the imaginary part of the quasinormal modes in the eikonal limit, assuming that the WKB approximation can be used\cite{1985ApJ291L33S,Konoplya:2019hlu}. In that limit, defining $\omega=\omega_R-i\omega_I$  as the frequency of the quasinormal modes, 
we have  $\omega_I=(n+\frac{1}{2})\lambda $ with $n$ the overtone number\cite{Cardoso2009}. Therefore, for those quasinormal modes we obtain the upper bound 
\begin{equation}
\omega_I r_H\leq \frac{1}{3\sqrt{3}}\left(n+\frac{1}{2}\right).
\end{equation} 
An associated bound for the real $\omega_R$ was derived in \cite{Hod:2013jhd}. However, this correspondence between null geodesics and quasinormal modes should be taken with caution. First, as shown by Konoplya in \cite{Konoplya:2017wot}, this correspondence can only be guaranteed in spacetimes where the effective potential associated to the quasinormal modes has a single peak, decaying at the event horizon and infinity, and it is limited to test fields
 and not for gravitational and other non-minimally coupled field. As shown in \cite{Konoplya:2017wot,Konoplya:2022gjp,Bolokhov:2023dxq}, this might not be the case in many alternative theories or even in asymptotically flat spacetimes. Additionally, the issue may be that the WKB approximation is valid but does not produce the complete spectrum of quasinormal modes\cite{Konoplya:2022gjp}. Therefore, the previous result should be considered valid only for those modes where the correspondence can be guaranteed, but it should not be taken as universal. As discussed in \cite{Konoplya:2017wot}, the same applies to the bounds for the real part of these modes as those found in \cite{Gallo:2015bda,Hod:2013jhd}.

Referring back to the bound \eqref{eq:lyapuni}, it is also related to a conjecture of \cite{Giataganas:2024hil} relating the Lyapunov exponent to the black hole entropy $S$, which reads \cite{Giataganas:2024hil}
\begin{equation}\label{eq:lambdaS}
\lambda^2 \leq \frac{\pi}{4S}.
\end{equation} {The authors' motivation for this empirical formula was to establish a new inequality that remains valid for a larger family of black holes. In particular, they observed that the inequality $ \lambda \leq \kappa_H $ which holds for the Schwarzschild spacetime, can be rewritten as \eqref{eq:lambdaS} by noting that in the Schwarzschild case, the surface gravity is given by $ \kappa_H = \frac{1}{2r_H} $, and the entropy is expressed as 
\begin{equation}
  S = \frac{A(r_H)}{4} = \pi r_H^2.  
\end{equation}
} 
This allowed them to explore its validity in other spacetimes.  
As a result, they showed that \eqref{eq:lambdaS} holds for the whole Kerr-Newman family of black holes.}

In contrast, from Eq.\eqref{eq:lyapuni}, we obtain
\begin{equation}\label{eq:lambdas2}
    \lambda^2 \leq \frac{1}{3r^2_H} = \frac{\pi}{3S},
\end{equation}    
which is a weaker bound than \eqref{eq:lambdaS}.
The inequality \eqref{eq:lambdas2} holds for a broader family of spherically symmetric black holes than  \eqref {eq:lambdaS}  of \cite{Giataganas:2024hil}. 
Since \eqref{eq:lambdas2} is a weaker bound, it is clear that it remains valid for the spacetimes studied in \cite{Giataganas:2024hil}. Naturally, for spherically symmetric black holes, writing the upper bound for $\lambda$ in terms of $r_H$ or $S$ is entirely equivalent. However, this equivalence does not hold for spinning black holes. It would be intriguing to establish a general proof of the validity of some form of these inequalities for rotating black holes that satisfy the Einstein equations and the dominant energy condition. Alternatively, it would be valuable to identify counterexamples where one or the other of these proposed inequalities is violated.

Furthermore, the bound \eqref{eq:lambdas2} can be improved if we additionally assume that $p_r$  is such that $|r^3p_r|$ decreases monotonically.
 In that case, it was proved by Hod \cite{Hod:2020pim} that $r_\gamma > \frac{3}{2}r_H$, leading to  
\begin{equation}
\lambda \leq \frac{\sqrt{\mu}}{r}\bigg|_{r_\gamma} \leq \frac{1}{\sqrt{3}r_\gamma} \leq \frac{2}{3\sqrt{3}r_H}.       
\end{equation}
Thus, for this family of black holes we find the universal upper bound $\lambda r_H \leq \frac{2}{3\sqrt{3}}$. This inequality also implies that for this class the relation  \begin{equation}\label{eq:lambdas3}
    \lambda^2 \leq \frac{4\pi}{27S},
\end{equation}    
is satisfied, which is sharper than the bound conjectured in \cite{Giataganas:2024hil} and also to eq.\eqref{eq:lambdas2}. 
The inequality \eqref{eq:lambdas3} is saturated by a Schwarzschild black hole.

{To conclude this section, we would like to relate some of the bounds found here to an alternative approach for studying circular null geodesics and their stability. This method, developed in \cite{Cunha:2022nyw} (see also \cite{Qiao:2022hfv,Qiao:2022jlu}), is based on a two-dimensional Riemannian optical metric\cite{Gordon:1923qva,Gibbons:2008rj,Crisnejo:2018uyn}, which has the property of sharing the spatial orbits of null geodesics of the spacetime as its own geodesics. More precisely, for a static and spherically symmetric spacetime of the form
\begin{equation}
ds^2 = -f(r)dt^2 + \frac{dr^2}{h(r)} + r^2 d\Omega^2,
\end{equation}
one assigns an optical metric:
\begin{equation}
dt^2 = \frac{1}{f(r)}\left(\frac{dr^2}{h(r)} + r^2 d\Omega^2\right).
\end{equation}
}
The stability of circular null geodesics at $r_\gamma$ follows from the sign of the Gaussian curvature $\mathcal{K}$ associated with these orbits, which at the position $r_\gamma$ is given by\cite{Cunha:2022nyw}
\begin{equation}
    \mathcal{K}(r_\gamma) = \frac{h(r_\gamma)}{2}\left(f''(r_\gamma) - \frac{f'(r_\gamma)}{r_\gamma}\right).
\end{equation}
If $\mathcal{K} < 0$, the orbits are unstable. 
After straightforward calculations, it can be verified using the expression for \eqref{lambda} for the Lyapunov exponent, and $N(r_\gamma) = 0$, that
\begin{equation}\label{eq:gausslambda}
\mathcal{K}(r_\gamma) = -\lambda^2.
\end{equation}

This relation can be easily understood by taking into account that the geodesic deviation equation in optical space for the deviation vector field $y^a$ along the circular null geodesic at $r_\gamma$ is a system of second order linear differential equations which linearly depend on the Gaussian curvature\cite{Casetti_1996,book:4378218}. 

Thus, taking into account the inequality \eqref{eq:lyapuni}, we can observe that at the position $r_\gamma$ of the innermost unstable null circular orbit of a black hole with event horizon at $r_H$, the Gaussian curvature of the associated optical metric is bounded from below and above as
\begin{equation}
-\frac{1}{3r_H^2} \leq \mathcal{K} < 0,
\end{equation}
that is, we have found that the optical space cannot have arbitrarily negative curvature at the unstable photon sphere.

\section{Summary}\label{sec:FIN}

{We} have proven that for every static and spherically symmetric black hole satisfying the Einstein equations and the dominant energy condition, there exists a set of universal bounds for the Lyapunov exponent $\lambda$ associated to unstable circular null geodesics in terms of the generalized surface gravity $\kappa_\gamma$ Eq.\eqref{eq:lambda20} and \eqref{eq:lambdar+}; the Unruh temperature Eq.\eqref{eq:nueva}; and the shadow size $R_{sh}$ Eq.\eqref{eq:lambdars}. Furthermore, we have reformulated some of these bounds in terms of the black hole's entropy Eq.\eqref{eq:lambdas2}.  { However, to our knowledge, a justification for the validity (or invalidity) of inequalities like \eqref{eq:lambdas2} based on fundamental physical principles remains unknown. Exploring possible underlying reasons could be an interesting direction for future research.}

For specific black holes where $|r^3 p_r|$ decreases
monotonically, we derived a sharper inequality Eq.\eqref{eq:lambdas3}. As a simple application, we show that these bounds also impose a straightforward constraint on the imaginary part of the quasinormal mode in the regime where the correspondence between circular null geodesics and quasinormal modes is valid.

It would be interesting to know whether some of the bounds proven here have analogs in other spherically symmetric black holes that do not satisfy the assumptions made in this work, specifically regarding the energy conditions or the requirement of being solutions to the Einstein equations. For instance, a generalized version of \eqref{eq:hod2} was proven in \cite{Gallo:2015bda}  for the $n$-dimensional Einstein-Gauss-Bonnet theory (see also \cite{Ghosh:2023kge}), and in \cite{Cvetic:2016bxi} Cvetic, Gibbons and Pope have shown that black holes arising from other gravitational theories also satisfy the bounds found in\cite{Gallo:2015bda}. Since the Lyapunov bound is, in some sense, related to the upper bound on the location of the photon sphere, it is hoped that more general versions of the bound for $\lambda$ exist for broader classes of metrics or alternative theories. More importantly, it would be desirable to test these inequalities for rotating black holes or to prove their validity in those spacetimes. However, in general, the Einstein equations for rotating spacetimes are more complex, and the null geodesics are typically non-planar. Therefore, further work and potentially different techniques will be required to address this problem.

\subsubsection*{Acknowledgments}

E.G. acknowledges financial support from CONICET and SeCyT-UNC. We also thank R. Konoplya for bringing attention to relevant works on the WKB approximation to quasinormal modes. E.G. thanks Frisaldo Hames, Dario Gallardo Mozart, and Guillermo Correas Diaz for motivating discussions, while T.M. appreciates discussions with Olivera Mi\v skovi\'c, Radouane Gannouji, and Rodrigo Olea.


\end{document}